\documentstyle[aps,epsfig]{revtex}
\def\ra{\rightarrow}

\def\be{\begin{equation}}
\def\ee{\end{equation}}
\def\bea{\begin{eqnarray}}
\def\eea{\end{eqnarray}}

\begin{document}
\title{Low energy atmospheric muon neutrinos in MACRO}
\author{{\bf The MACRO Collaboration:}\\
M.~Ambrosio$^{12}$, 
R.~Antolini$^{7}$, 
G.~Auriemma$^{14,a}$, 
D.~Bakari$^{2,17}$,
A.~Baldini$^{13}$, 
G.~C.~Barbarino$^{12}$, 
B.~C.~Barish$^{4}$, 
G.~Battistoni$^{6,b}$, 
R.~Bellotti$^{1}$, 
C.~Bemporad$^{13}$
% ...
\\}
\address{
1. Dipartimento di Fisica dell'Universit\`a di Bari and INFN, 70126 
Bari,  Italy \\
2. Dipartimento di Fisica dell'Universit\`a di Bologna and INFN, 
 40126 Bologna, Italy \\
3. Physics Department, Boston University, Boston, MA 02215, 
USA \\
4. California Institute of Technology, Pasadena, CA 91125, 
USA \\
5. Department of Physics, Drexel University, Philadelphia, 
PA 19104, USA \\
6. Laboratori Nazionali di Frascati dell'INFN, 00044 Frascati (Roma), 
Italy \\
7. Laboratori Nazionali del Gran Sasso dell'INFN, 67010 Assergi 
(L'Aquila),  Italy \\
8. Depts. of Physics and of Astronomy, Indiana University, 
Bloomington, IN 47405, USA \\
9. Dipartimento di Fisica dell'Universit\`a dell'Aquila  and INFN, 
 67100 L'Aquila,  Italy \\
10. Dipartimento di Fisica dell'Universit\`a di Lecce and INFN, 
 73100 Lecce,  Italy \\
11. Department of Physics, University of Michigan, Ann Arbor, 
MI 48109, USA \\	
12. Dipartimento di Fisica dell'Universit\`a di Napoli and INFN, 
 80125 Napoli,  Italy \\	
13. Dipartimento di Fisica dell'Universit\`a di Pisa and INFN, 
56010 Pisa,  Italy \\	
14. Dipartimento di Fisica dell'Universit\`a di Roma ``La Sapienza" and INFN, 
 00185 Roma,   Italy \\ 	
15. Physics Department, Texas A\&M University, College Station, 
TX 77843, USA \\	
16. Dipartimento di Fisica Sperimentale dell'Universit\`a di Torino and INFN,
 10125 Torino,  Italy \\	
17. L.P.T.P., Faculty of Sciences, University Mohamed I, B.P. 524 Oujda, Morocco \\
18. Dipartimento di Fisica dell'Universit\`a di Roma Tre and INFN Sezione Roma Tre, 
 00146 Roma,   Italy \\ 	
\vspace {0.2cm}
{\footnotesize
$a$ Also Universit\`a della Basilicata, 85100 Potenza,  Italy -
$b$ Also INFN Milano, 20133 Milano, Italy\\
$c$ Also Istituto TESRE/CNR, 40129 Bologna, Italy -
$d$ Also Scuola Normale Superiore di Pisa, 56010 Pisa, Italy\\
$e$ Also Universit\`a di Trieste and INFN, 34100 Trieste, Italy -
$f$ Also Dip. di Energetica, Universit\`a di Roma,  00185 Roma,  Italy \\
$g$ Also Institute for Nuclear Research, Russian Academy
of Science, 117312 Moscow, Russia \\
$h$ Also Institute for Space Sciences, 76900 Bucharest, Romania -
$i$ The Colorado College, Colorado Springs, CO 80903, USA\\
$l$ Also Dept. of Physics, Pennsylvania State University, 
    University Park, PA 16801, USA\\
$m$ Also Dipartimento di Fisica dell'Universit\`a della Calabria, Rende (Cs), Italy \\
$n$ Also Dept. of Phys. James Madison University, Harrisonburg, VA 22807, USA\\
$*$ Corresponding author. e-mail: SPURIO@BO.INFN.IT 
}}
\maketitle
\begin {abstract}
We present the measurement of two event samples induced by atmospheric
$\nu_\mu$
of average energy $ \overline {E}_\nu \sim 4 \ GeV$.
In the first sample, the neutrino interacts inside the MACRO detector
producing an upward-going muon leaving the apparatus.
The ratio of the number of observed to expected events is
$ 0.57 \pm0.05_{stat} \pm0.06_{syst} \pm0.14_{theor}$ with an
angular distribution similar to that expected from the Bartol
atmospheric neutrino flux.
The second is a mixed sample of internally produced downward-going
muons and externally
produced upward-going muons stopping inside the detector.
These two subsamples are selected by topological criteria; the lack
of timing information makes it impossible to distinguish stopping from
downgoing muons.
The ratio of the number of observed to expected events is
$0.71 \pm 0.05_{stat} \pm0.07_{syst} \pm0.18_{theor}$ .
Using the ratio of the two subsamples 
(for which most theoretical  uncertainties cancel) we can test the pathlength
dependence of the oscillation hypothesis. The probability of agreement
with the no-oscillation hypothesis is $5\%$ .

The deviations of our observations from the expectations has a preferred
interpretation in terms of $\nu_\mu$ oscillations
with maximal mixing and $\Delta m^2 \sim 10^{-3} \div 10^{-2}\ eV^2$.
These parameters are in agreement with our results from upward
throughgoing muons, induced by $\nu_\mu$ of much
higher energies.
\end {abstract}
\nopagebreak
\twocolumn
\narrowtext
\setcounter{equation}{0}

The results from several underground detectors which measure the
flux of atmospheric neutrinos give strong indication that $\nu_\mu$'s
oscillate into neutrinos of another type
\cite{SUPERK98,SOUDAN99,MACRO98}.
Fully-contained and partially-contained neutrino-induced events observed in
underground detectors come from neutrinos of energy $\sim 1 \ GeV$.
The flux of atmospheric neutrinos of several tens of $GeV$ can be inferred
from the measurement of neutrino-induced upward-going muons that traverse
the entire detector ({\it up-throughgoing muons)}.
The hypothesis of neutrino oscillations, with best-fit parameters
$\sin^2 2\theta_{mix} \sim 1$ and $\Delta m^2$ in the range of a few times
10$^{-3}$ eV$^2$, can explain the observed anomalies both in the
ratio of contained $\nu_\mu$ to $\nu_e$ events (Super-K, Soudan 2)
and in the zenith angle
distribution of up-throughgoing muons (MACRO, Super-K). 

The MACRO detector measures both the high energy
(median energy $\sim 50 \ GeV$) and few $GeV$ energy atmospheric
neutrino fluxes.
In Ref. \cite{MACRO98,MACRO95} the interpretation of the data in terms of
$\nu_\mu$ oscillations came from a deficit
and from an anomalous zenith angle distribution of the observed
up-throughgoing muons originating from $\nu_\mu$ interactions in the
rock below the detector.
Here we report on the measurement of the flux of lower energy
$(\overline E_\nu \sim 4\ GeV$) atmospheric neutrinos through the
detection of $\nu_\mu$ interactions inside the apparatus
(yielding partially contained upgoing and downgoing muons)
and by the detection of externally produced upward-going muons
stopping inside the detector \cite{LOWNU}.

MACRO \cite{MACRO93} is a large area, modular tracking
detector located in Hall B of the Gran Sasso Underground Laboratory in
Italy,
with an average rock overburden of 3700 hg/cm$^2$.
It is a rectangular box, 76.6~m~$\times$~12~m~$\times$~9.3~m,
divided longitudinally into six supermodules and vertically into
a lower part (4.8 m high) and an upper part (4.5 m high).
The active detection elements are planes of streamer
tubes for tracking and liquid scintillation counters for fast timing.
The lower half of the detector is filled with streamer tube planes
alternating with trays of crushed rock, which provide most of the
5.3 $kton$ target
mass for partially-contained neutrino interactions. The upper part is
hollow and contains the electronics racks and work areas. There are 10
horizontal streamer tube planes in the bottom half of the detector, and 4
planes on the top, all with wire and 27$^\circ$ stereo strip readouts.
Six vertical planes of streamer tubes cover each side of the detector.
The intrinsic angular resolution for muons is between
0.1$^\circ$ and 1.0$^\circ$ depending on the track length.
The scintillator system consists
of three widely-separated layers of horizontal boxes,
and on each vertical side of the detector a layer of vertical boxes
inserted between the streamer tubes.
The time (position) resolution for muons in a scintillator
box is about 500~ps ($\sim 11$~cm). The direction of the muons
passing across MACRO is determined by the time-of-flight between two
layers of scintillation counters.

The results presented in this letter come from 4.1 live years of data 
taking with the full detector, from April 1994 to February 1999.

\begin{figure}
\begin{center}
\epsfig{file=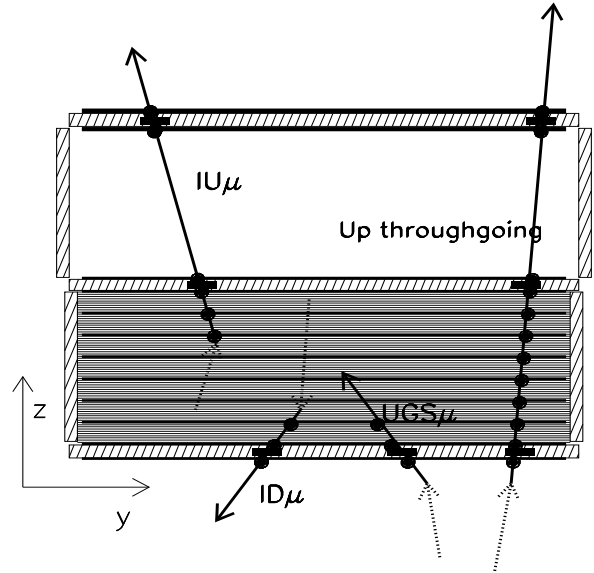,height=7.0cm,width=8cm}
\caption{\label{fig:topo}\small Event topologies induced by neutrino
interactions in or around MACRO.
$IU$= Internal Upgoing $\mu$;
$ID$= Internal Downgoing $\mu$;
$UGS$= Upgoing Stopping $\mu$;
Up throughgoing = upward throughgoing $\mu$.
The the black circles indicate the streamer tube hits, and the black
boxes the scintillator hits. The $T.o.F.$ of the muon is measured
for the $IU$ and Up throughgoing events.}
\end{center}
\end{figure}

\vspace{-10pt}

\noindent About $33\times 10^6$ downgoing muons were collected,
and were used to monitor the detector efficiency, the running
conditions and the acceptance.
The trigger rate due to downgoing muons is $0.3\ Hz$.
The trigger efficiency for each scintillation counter and for the
streamer tubes was monitored over the data taking
period using the downgoing muons.

Two samples of atmospheric muon neutrinos in the few-$GeV$ energy range
are measured. In the first sample
(up partially-contained or $IU$=Internal Upgoing $\mu$ events)
there are (mainly) events induced by
charged current (CC) interactions of upgoing $\nu_\mu$
inside the lower part of MACRO.
An upgoing muon is produced, which
crosses two scintillation layers (Fig. \ref{fig:topo}),
so that the measurement of the direction is made through time-of-flight.

\begin{figure}
\epsfig{figure=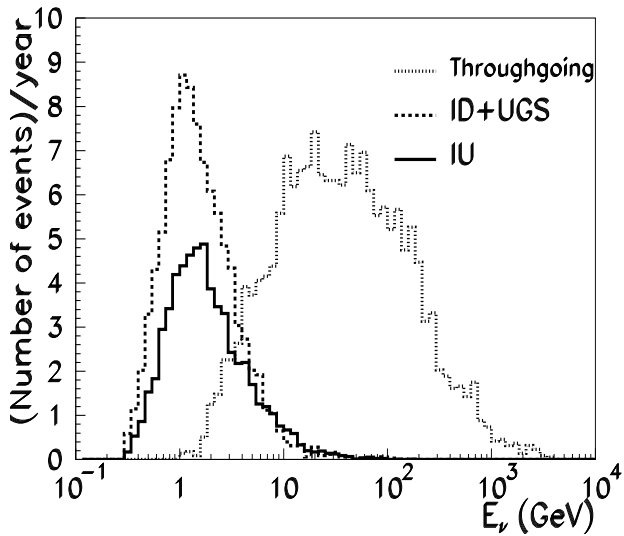,height=7.2cm,width=8cm}
\caption {\label{fig:entop}\small
Monte Carlo simulated distribution of the
parent neutrino energy giving rise to the three different
topologies of events detectable by MACRO. The distributions are
normalized
to one year of data taking; the analysis cuts are included.}
\end{figure}

The second sample is a mix of upgoing and downgoing events.
The partially contained downgoing events
(down partially-contained or $ID$=Internal Downgoing $\mu$)
are induced by downgoing $\nu_\mu$, interacting in the lower part of
MACRO.
The upgoing stopping events ( $UGS$ = Upward Going Stopping muons)
are induced by interactions of upgoing $\nu_\mu$ below the detector
yielding an upgoing muon which stops inside the detector.
Both the down partially-contained and the upgoing stop events
cross only the bottom layer of liquid scintillation counters
(see Fig. \ref{fig:topo})
and are identified by means of topological criteria.
The lack of timing information makes it impossible to
distinguish between the two subsamples.
Fig. \ref{fig:entop} shows the parent neutrino energy distribution
from a Monte Carlo calculation for the
three event topologies detectable in MACRO.
The energy spectrum and the median energy of
the two samples presented in this letter are almost the same.

\vspace{-6pt}

\begin{figure}
\begin{center}
\epsfig{file=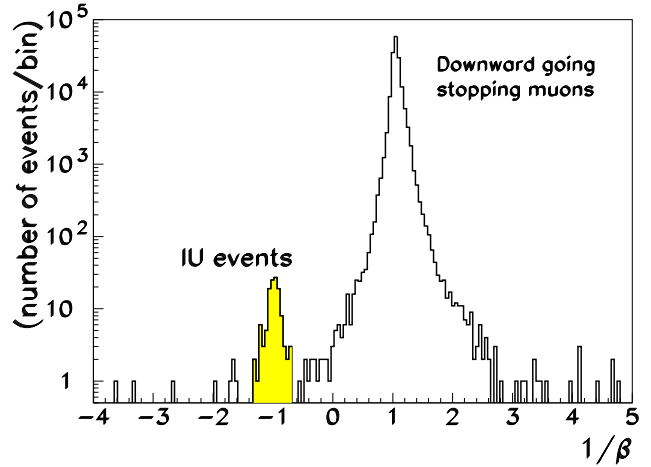,height=6cm,width=8cm}
\vspace{5pt}
\caption{\label{fig:sbetatk}\small
The $1/\beta$ distribution of partially-contained events after all software
cuts. 121 events are in the $IU$ event signal region,
$-1.3 < 1/\beta < -0.7$, with 5 estimated background events.}
\end{center}
\end{figure}

The identification of $IU$ events is based on topological
criteria and time-of-flight measurements.
The main requirement is the presence of at least two hit
scintillator clusters, respectively in the center
layer and in the upper part of the apparatus
(see Fig. \ref{fig:topo}).
This is the expected topology for a neutrino interacting in the
lower detector and producing an upward-going muon with enough energy to
exit the apparatus. It is also the topology of the much more numerous
downgoing muons stopping in the lower detector.
Scintillation timing allows the separation of the two classes of events.
Moreover, the scintillators are required to match a streamer tube track
reconstructed in space by our standard track-finding
algorithms \cite{MACRO93}.
For $IU$ candidates, the lowest point of a track (the starting point)
must be inside the apparatus as a condition for the containment of the
$\nu_\mu$ interaction vertex. To reject fake partially-contained events
entering from a detector insensitive zone, 
the extrapolation of the track in the lower part of
the detector must geometrically cross and not fire
at least one scintillator layer and one streamer
tube plane, or at least three planes of streamer tubes.
These conditions were tuned on Monte Carlo simulated events,
including evaluation of detector inefficiencies.
Other cuts are applied to reject background events from downgoing atmospheric
muons. They are related to the goodness of the geometrical agreement
between scintillator hits and the streamer track, to the proper operation of
the scintillation counters and to the quality of the time measurement.
The measured $1/\beta$ distribution after all analysis cuts (including
the requirements of vertex containment) is shown in Fig.
\ref{fig:sbetatk}.
The measured
muon velocity $\beta c$ is evaluated with the sign convention that
upgoing muons have $1/\beta \sim -1$.
A total of 121 events survive in the range $-1.3 < 1/\beta < -0.7$, which
is taken as the range of $IU$ signal.

From the time distribution of Fig. \ref{fig:sbetatk} one expects some
background events; they are
mostly due to wrong time measurements or secondary particle
hits, yielding an almost flat $1/\beta$ distribution.
The fit of the distribution in the range $-6.0 < 1/\beta < -0.3$ 
to a gaussian plus a straight line gives an estimated background of
5 events in the signal region.
After background subtraction, we have 116 up partially-contained events.

\vspace{-8pt}

The identification of $ID+UGS$ events is based on topological
criteria. The candidates have a track starting (ending) in the lower
apparatus, and crossing the bottom detector face.
The track must also be located or oriented in such a way that it could not
have entered or exited undetected through insensitive zones in the apparatus.
Events with scintillator hits outside the bottom layer, or
with the reconstructed track pointing to a detector insensitive zone
between modules, are rejected.
The standard muon tracking procedure \cite{MACRO93} is based on
at least four aligned hits in the
streamer tubes. This corresponds to a minimum traversed detector thickness
$t_v \sim 200\ g/cm^2$ {\it (standard sample)}.
A dedicated retracking procedure was applied to all remaining
events. The retracking procedure
requires at least three streamer tube hits
($t_v \sim 100\ g/cm^2$), aligned with respect to a fired scintillation
counter in the bottom layer.
The number of retracked events is $\sim 5\times 10^5$
(to be compared with the $\sim 32\times 10^6$ standard muon tracks); these
events follow the same analysis as the standard sample.

After the software cuts, 879 events survive.
Some of them are tracked incorrectly (mostly by the retracking),
or are bending downgoing muons, entering from a detector insensitive zone.
Due to the bending, only a fraction of the streamer tube hits are
used by the tracking algorithms.
In order to reject these fake candidates, we made
a double scan with the MACRO Event Display\cite{MACRO93}. To
eliminate any bias from the scan procedure, and to evaluate the
absolute and relative scanning efficiencies, Monte Carlo (MC)
simulated events (described below) were randomly injected into the
data sample before the scanning stage. Two physicists independently
scanned the merged sample. At the end of the scan,
200 events in the real data (106 $ID+UGS$ candidates are retracked events)
are accepted as upgoing stopping or partially contained
downgoing muons. 97\% of the real events selected 
by one physicist were also
selected by the other.

Downgoing muons which pass near
or through the detector may produce low-energy, upgoing particles,
which could simulate neutrino-induced
upgoing muons if the downgoing muon misses the detector.
This background has been evaluated using a full
simulation, based on our measurements \cite{macrouppi}.
The background is $7\pm 2$ events. A second background source could
arise from atmospheric muons and detector inefficiency. Using
a simulated sample of $10^7$ atmospheric muons,
which includes measured detector inefficiencies,
no events were selected by the above described procedure.
After background subtraction, 193 events represent the down partially-contained
plus upgoing stopping signal.

The expected number of neutrino-induced events was estimated from GMACRO
\cite{MACRO93}, a GEANT-based \cite{Brun87} full MC detector
simulation.
The $\nu_e$ and $\nu_\mu$ interaction rates have been
computed using the atmospheric neutrino flux of the Bartol group
\cite{Agrawal96} and the neutrino cross sections of Ref. \cite{Lipari95}.
In this cross-section model,
the contributions of the exclusive channels of
lowest multiplicity (quasi-elastic and single pion production)
are calculated separately from deep
inelastic scattering (DIS). The DIS contribution
to the $\nu$N cross section was computed using the GRV-LO-94
\cite{Gluck95} parton distribution functions.
Using these neutrino fluxes and cross sections for
$E_\nu \ge 300\ MeV$, we expect a total ($CC+NC$) interaction rate of
$71.5\ / kton\cdot y\ (\nu_e +\overline \nu_e)$ and
$148.1\ / kton\cdot y\ (\nu_\mu +\overline \nu_\mu)$.
Two simulated samples have been generated, because of the different
vertex locations for the $IU$ and the $ID+UGS$ events;
the simulated events were processed with the same analysis
chain as the data.

For the $IU$ events, a sample of $10^5$
interactions inside the apparatus was
generated (equivalent to $85.9$ years live time).
The simulation indicates (see Table 1) that $87 \%$ of detected
$IU$ events come from $\nu_\mu$ charged current (CC) interactions,
$9 \%$ from $\nu_e$ CC and the remaining fraction from neutral current
(NC) interactions.
Due to detector inefficiencies and analysis algorithm failures, some
neutrino-induced events originating in the rock surrounding the detector are
expected
to contribute to the selected sample of up partially-contained events
(upward-throughgoing $\mu$'s appearing as partially-contained). The
vertex containment requirements reduce this background
to about $1\%$, evaluated using a simulated sample of up throughgoing
muons \cite{MACRO98}.
The fully-automated selection gives a total number of
202 expected events in the $IU$ event signal region,
$-1.3 < 1/\beta < -0.7$ for $4.1\ y$ of live-time .

For the $ID+UGS$ events, $1.16\times 10^6$ neutrino interactions were
simulated in a larger volume (including the experimental hall, the
detector and the surrounding rock). The generated events correspond
to a live time of $31.1\ y$.
The dimensions of the interaction volume (with $13\ m$ of rock below the
detector, and a total rock mass of $165\ kton$ plus $5.3\ kton$ of
the apparatus itself) were chosen to reduce
to less than $1\%$ the number of $\nu_\mu$-induced stopping muons
produced outside that volume.
The 2199 events which survived the software selection for the $ID+UGS$
were merged with the real events which passed the same software
selection, and visually scanned. After the scan procedure 2074
(=94.3\%) of the simulated and reconstructed events were accepted,
together with the 200 real events. The expected rate is
273 events in $4.1\ y$ live-time.
In Table \ref{tab:simu} we give
the main features of the $IU$ and $ID+UGS$ simulated samples
(expected rate, percentage of CC $\nu_\mu$ interactions,
median parent neutrino energy, and fraction of events induced by upgoing
neutrinos).

\begin{figure}
\begin{center}
\epsfig{file=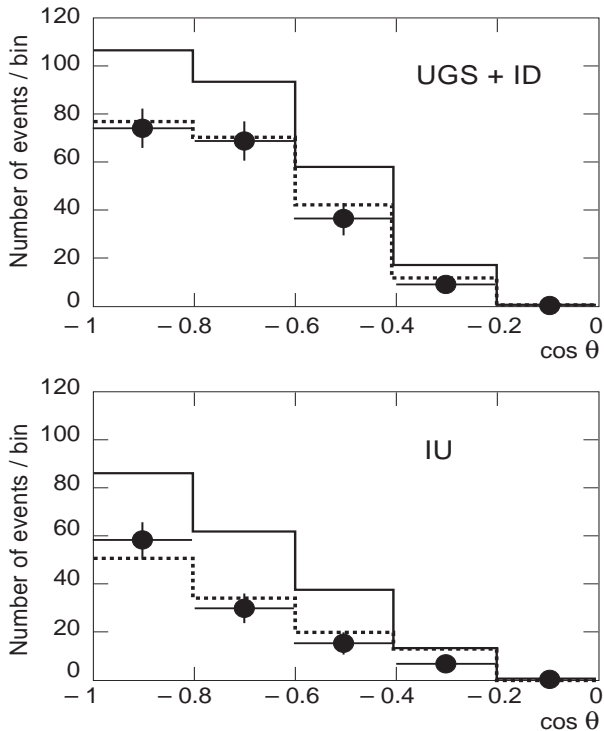,height=9.8cm,width=8cm}
\caption{\label{fig:cos}\small
Measured distributions in the cosine of the zenith angle $\Theta$ for
the a) $ID+UGS$ and b)$IU$ events (black points with error bars).
The solid lines are the Monte Carlo predictions assuming no
oscillations. The dashed lines are the
expectations for $\nu_\mu \rightarrow \nu_\tau$ oscillations with
$\Delta m^2 = 2.5 \times 10^{-3} eV^{2}$ and maximal mixing.}
\end{center}
\end{figure}

The number of detected $IU$ events in the
4.1 $y$ live-time is 116, while the expected number is 202.
The ratio of the measured to the expected events is
$R_{IU} = ({{Data}\over {MC}})_{IU}=
0.57 \pm 0.05_{stat} \pm 0.06_{syst} \pm 0.14_{theor}$.
For the $ID+UGS$ sample, 193 events are detected while 273 are expected.
The ratio is $R_{ID+UGS} =
0.71 \pm 0.05_{stat} \pm 0.07_{syst} \pm 0.18_{theor}$.
Each data set is (within errors) consistent with a constant
deficit (43\% for the $IU$ sample, 29\% for the ID+UGS)
in all zenith bins compared to the Monte Carlo expectations
assuming no oscillations.
Fig. \ref{fig:cos} shows the zenith angle distributions for the two
measured data sets and for the Monte Carlo simulations.
Due to the analisys cuts and to the apparatus acceptance, there are no
events in the last bin of the two distributions.
The expectations are affected by a systematic theoretical
error due to the uncertainties regarding
the atmospheric neutrino flux and the neutrino cross sections.
At present there is no unique and
reliable estimate of the total theoretical uncertainty for the
rate calculations. Each experimental group, and for each event category,
has its own way to estimate it.
For this analysis we conservatively
estimate 20\% for the flux and 15\% for the cross section, which
add in quadrature to an error of 25\%. For our high energy events
\cite{MACRO98} we quoted 17\%, while in the recent SuperKamiokande
\cite{skstop} analysis of neutrino-induced stopping muons 22\% was quoted .

Our measured value of $R_{IU} = 0.57$ is quite far from its expected
value of unity. If we ignored theoretical errors (i.e. if we
assumed the flux and cross section as we modeled them were 
accurate), an experiment with our statistical and experimental
uncertainties would only fluctuate so far from unity with
probability $2.5 \times 10^{-4}$. However, if we take the $25\%$
theoretical error into account, the probability becomes $6.5\%$.
$R_{ID+UGS} = 0.71$ also differs from unity, though not as
significantly as $R_{IU}$.

If the observed deficit were due only to an overall theoretical
overestimate of the neutrino flux or cross sections,
one would expect $R_{IU} \sim R_{ID+UGS}$ (small differences would
remain due to residual geomagnetic effects).
Furthermore, the theoretical uncertainties
largely cancel if the ratio ${{IU/ID+UGS}}$ 
between the measured number of events
is compared with the expectation.
The partial error cancellation arises from the almost identical energy
spectra of parent neutrinos for the two samples of events; we
evaluated the remaining error as 5\%.
The experimental systematic uncertainty for the ratio is estimated at 6\%.
The measured ratio is ${{IU}\over {ID+UGS} }= 0.60\pm 0.07_{stat}$, while
the expectation without oscillations is $0.74\pm 0.04_{sys} \pm
0.04_{theo}$.
The probability to obtain a ratio so far from the expected one is 5\%,
almost independent of the neutrino flux and neutrino cross sections used
for the predictions.
\vspace{-8pt}

We investigated if the observed discrepancies between data and expectations
could be explained by possible systematic effects.
The detector mass is known to $\pm 5\%$.
The uncertainty for the detector acceptance was estimated by comparing
the shape of the zenith distribution of downward-going muons stopping inside
the detector with a MC expectation based on the known rock
overburden: the two distributions agree within $6\%$.
Other uncertainties arise from the live-time estimate ($3\%$), the
effective
containment of the interaction vertex depending on the simulation of the
detector response to internal neutrino interactions ($4\%$) and the
background
subtraction ($4\%$). Adding all these contributions in quadrature yields
our quoted experimental systematic
uncertainty of $10 \%$, too small to account for the observed discrepancy.

\begin{figure}
\begin{center}
\epsfig{file=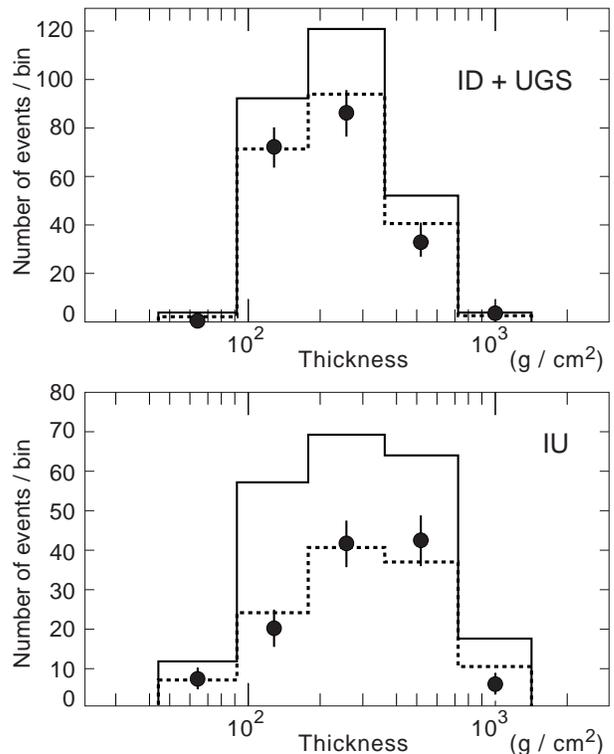,height=10cm,width=8cm}
\caption{\label{fig:emin}\small
Distribution of the amount of material ($t$ in $g\ cm^{-2})$ in the detector
traversed by
$IU$ and $ID+UGS$ events. The data points (black points with error bars) are
compared with the Monte Carlo prediction and no oscillations
(solid histogram).
The expectation for $\nu_\mu \rightarrow \nu_\tau$ oscillations with
maximal mixing and $\Delta m^2 = 2.5 \times 10^{-3} eV^{2}$ is shown by
the dashed histogram.}
\end{center}
\end{figure}
\vspace{-8pt}

The number of expected events was also evaluated using the NEUGEN
neutrino event generator \cite{NEUGEN} (developed by the Soudan and MINOS
collaborations) as input to our MC simulations.
The NEUGEN generator predicts $\sim 6\% \ (5\%)$ fewer $IU$ (ID+UGS)
events than our default generator \cite{Lipari95}, well
within the estimated systematic uncertainty for
neutrino cross sections ($\sim 15\%$).

Our data disfavor the no-oscillations hypothesis regardless
of overall normalization; they are consistent with neutrino
oscillations ($\nu_\mu$ disappearance) with maximal mixing and
$\Delta m^2 \sim (1 \div 20) \times 10^{-3}\ eV^2$. As a ``test point'',
we use the best-fit parameters from our high-energy analyses
\cite{MACRO98,MACRO99}, $\Delta m^2 =2.5\times10^{-3} \ eV^2$ and
$sin^2 2\theta_{mix}=1$.
The predicted numbers of events and the angular
distributions are indicated by the dashed histograms in Fig.
\ref{fig:cos}; they are in good agreement with the measured data.

For $\Delta m^2 \sim (1 \div 20) \times 10^{-3}\ eV^2$,
upgoing neutrinos
(which induce $IU$ and $UGS$ events),
which travel thousands of kilometers through the Earth, are
reduced by $50\%$.
Almost no reduction is expected for downgoing partially-contained muons.
In this scenario, and for a pure $\nu_\mu$ CC interaction sample,
the expected event rate is 1/2 of the $IU$ and
3/4 of the $ID+UGS$ predictions without oscillations.
The predicted reduction for upgoing $\nu_\mu$ is less than 1/2
because of the $\nu_e$ and NC event contaminations.
Our data disfavor $\Delta m^2 > 10^{-2}\ eV^2$, for which
the $ID$ events are also reduced; both
the $ID+UGS$ and $IU$ event rates are $\sim 1/2$ of the
no-oscillations expectation.
We also disfavor $\Delta m^2 < 10^{-3}\ eV^2$, for which the shape of the
angular distributions (Fig. \ref{fig:cos}) is modified.

Assuming oscillations (with the ``test point'' parameters) 115 up 
partially-contained
and 202 down partially-contained plus upgoing stopping
events are expected.
For the $IU$ events, the reduction from the no oscillations hypothesis
is 0.57, to be compared with the measured value of $R_{IU} = 0.57$.
For the $ID+UGS$ events, it
is 0.76, to be compared with $R_{ID+UGS} = 0.71$.
The quoted numbers use our default
normalization. Recent flux calculations \cite{fluka}
suggest that the Bartol flux which we use may be too high (though
within the quoted theoretical error). 
As far as the event rates are concerned,
a lower normalization of the flux can still be 
partially compensated at low energies by different interaction cross
sections for neutrinos.

The event distributions as a function of
the muon pathlength inside the detector have also been studied, as an
independent consistency check.
In Fig. \ref{fig:emin}
the data (black points) are compared with the MC expectation (solid lines
for no neutrino oscillations; dashed lines for
oscillations at our test point). The shapes are similar,
and the data prefer the reduced normalization of the oscillation
prediction.

\begin{figure}
\begin{center}
\mbox{
\epsfig{file=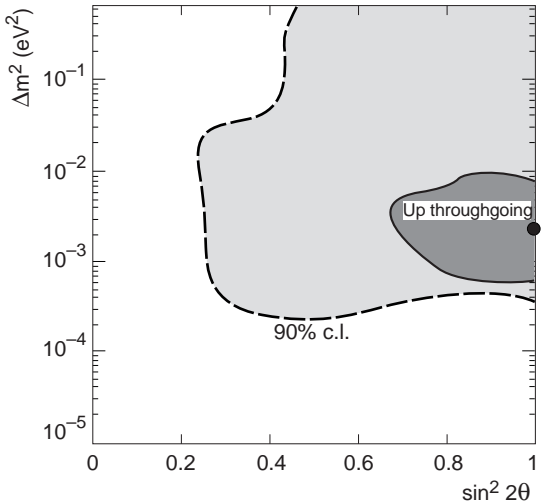,width=7.3cm,height=7.3cm} }
\caption{\label{fig:exclu} Allowed contours at
90\% C.L. for $\nu_\mu \rightarrow \nu_\tau$ oscillations 
obtained by combining the low energy neutrino events ($IU$
and $ID+UGS$) using the prescription of \protect\cite{feldman}.
The 90\% C.L. contour and the best fit point from our high-energy
analysis \protect\cite{MACRO99} is also shown as a dashed line.}
\end{center}
\end{figure}
\vspace{-16pt}

We estimated the most likely values of $\Delta m^2$ and $sin^2 2\theta_{mix}$
using a $\chi^2$ method for data and Monte Carlo for
the data of Fig. \ref{fig:cos}.
The $\chi^2$ was computed with ten degrees of freedom:
the histograms ($2\times 4$ bins, normalized so as to
contain only distribution shape information),
the ${{IU}\over {ID+UGS}}$ ratio and the
overall normalization.
The statistical and systematic errors are added in quadrature;
the systematic uncertainty is 10\%
in each bin of the angular distributions, 5\% for the
ratio, and 25\% for the normalization.
Fig. \ref{fig:exclu} shows the 90\% confidence level region,
based on the application of the MC prescriptions of Ref. \cite{feldman}
on a $(sin^2 2\theta_{mix}, \Delta m^2)$ grid.
The expected flux $\Phi^i_{osc}$ for a given point of
$(sin^2 2\theta_{mix}, \Delta m^2)$ in the grid is obtained
by weighting each
simulated event with its surviving probability
$P(\nu_\mu \ra \nu_\mu)
= 1-sin^22\theta_{mix} \cdot sin^2(1.27 \cdot \Delta m^2 L/E_\nu)$
in that bin.
The maximum of the $\chi^2$ probability
(97\%) occurs at $sin^22\theta_{mix} =1.0$;
this $\chi^2$ probability is almost constant in the interval
$\Delta m^2 = (1 \div 20)\times 10^{-3}\ eV^2$.
In the region of the maximum, the
flux normalization factor is $1.02$ in both data
sets ( i.e. the data are 2\%
higher than the oscillated predictions with our normalization).

In conclusion, we presented
measurements of two samples of events induced by relatively
low-energy neutrinos ($\overline E_\nu \sim 4\ GeV$)
interacting in MACRO or in the surrounding rock.
The neutrinos originate from cosmic ray
interactions in the upper atmosphere above the detector
(downgoing events) or up to $\sim 13000\ km$
away on the opposite side of the Earth (upgoing events).
The ratio of the number
of observed to expected events (no oscillations) is
$0.57 \pm 0.05_{stat} \pm 0.06_{syst} \pm 0.14_{theor}$ for the $IU$ events
and $0.71 \pm 0.05_{stat} \pm 0.07_{sys} \pm 0.18_{theo}$ for the $ID+UGS$.
Within statistics, the observed deficits are uniform over the zenith angle.
From the ratio of $IU$ to $ID+UGS$, the probability that there is an
overall reduction in the number of neutrino-induced muons is 5\%.
The hypothesis of muon neutrino oscillations
explains the different deficits
in $IU$ and $ID+UGS$ events with higher probability.
The large theoretical uncertainties for the neutrino flux and
cross sections is dominant in each data set; the ratio of the two
low energy samples is dominated by statistical uncertainties.
The regions with
$\Delta m^2> 3\times 10^{-4}\ eV^2$ and $sin^22\theta_{mix}> 0.25$
are allowed at $90\%$ C.L. The best region corresponds to
maximal mixing and $\Delta m^2 = (1 \div 20) \times 10^{-3}\ eV^2$.
This result confirms the
scenario proposed by the measurement of
higher-energy neutrino-induced muons by MACRO
\cite{MACRO98,MACRO99},
as well as by other experiments \cite{SUPERK98,SOUDAN99}, all of which favor
the $\nu_\mu$ oscillation
hypothesis with maximal mixing and
$\Delta m^2$ of a few times $10^{-3}\ eV^{2}$.

\begin{table}
\begin{center}
\begin{tabular}
{ccccc}
& Rate & $\nu_\mu$ CC & $\overline E_\nu$ & $\nu \uparrow$ \\
&($ev/y$)& (\%) & (GeV) & (\%) \\ \hline
IU & 49.3 & 87 & 4.2 & 94 \\
ID+UGS& 66.7 & 87 & 3.5 & 51 \\
\end{tabular}
\end {center}
\caption {\small Summary of the properties of simulated $IU$ and $ID+UGS$
events.
From column 2 to 5: expected event rate (no neutrino oscillations);
percentages of events induced by $\nu_\mu$ CC interactions;
median parent neutrino energies (5\% of events with the highest/lowest
energies were excluded); percentages of events induced by upgoing
neutrinos.}
\label{tab:simu}
\end{table}

\vskip 0.2 cm
We acknowledge the support of the staff of
the Gran Sasso Laboratory and of
the Institutions participating in the experiment. We
thank the Istituto Nazionale di Fisica Nucleare (INFN), the U.S.
Department of
Energy and the U.S. National Science Foundation for their support.
We thank INFN, FAI, ICTP (Trieste), NATO and WorldLab for providing
fellowships and grants for non-Italian citizens.


\begin{thebibliography}{99}

\bibitem{SUPERK98}
SuperKamiokande Collaboration, Y. Fukuda {\it et al.}, Phys. Rev. Lett.
{\bf 81} (1998) 1562.
%
\bibitem {SOUDAN99}
Soudan 2 Collaboration, W.W.M. Allison {\it et al.}, Phys. Lett. {\bf
B449}, (1999) 137.
%
\bibitem{MACRO98}
MACRO Collaboration, M. Ambrosio {\it et al.}, Phys. Lett. {\bf B434}
(1998)451.
%
\bibitem{MACRO95} MACRO Collaboration, M.Ambrosio {\it et al},
Phys. Lett. {\bf B357} (1995) 481.
%
\bibitem{LOWNU}
Preliminary results were presented at the conferences/workshops of:
Alcal\`a de Henares 1998, hep-ex/9808001; Vulcano 1998, hep-ex/9809003;
Takayama 1998, hep-ex/9810008; DPF 1999, hep-ex/9903030; La Thuile 1999,
hep-ex/9906019; ICRC 1999, INFN/AE-99/10(1999).
%
\bibitem{MACRO93}
MACRO Collaboration, S. Ahlen {\it et al.}, Nucl. Instr. and Meth. {\bf
A324} (1993) 337.
%
\bibitem{macrouppi} MACRO Collaboration, M. Ambrosio {\it et al.},
Astroparticle Physics 9 (1998) 105
%
\bibitem{Brun87}
R. Brun {\it et al.}, CERN report DD/EE84-1 (1987).
%
\bibitem{Agrawal96}
V. Agrawal {\it et al.}, Phys. Rev. {\bf D53} (1996) 1314.
%
\bibitem{Lipari95}
P. Lipari {\it et al.}, Phys. Rev. Lett.{\bf 74} (1995) 4384.
%
\bibitem{Gluck95}
M. Gl\"{u}ck {\it et al.}, Z. Phys. {\bf C67} (1995) 433.
%
\bibitem{skstop} SuperKamiokande Collaboration, Y.Fukuda {\it et al.},
{\it Neutrino-induced upward stopping muons in Super-Kamiokande},
hep-ex 9908049-v3 (1-Dec-1999).
%
\bibitem{NEUGEN}
Private communication from the MINOS collaboration; see also H.M.
Gallagher, {\it Neutrino Oscillation Searches with the Soudan 2 Detector},
Ph.D. thesis, University of Minnesota (1996).
%
\bibitem{MACRO99} MACRO Collaboration, ({\it Neutrino oscillations
at high energies by MACRO}), INFN/AE-99/09. hep-ex 9905025.
%
\bibitem{fluka} G. Battistoni et al., hep-ph 9907408, to be published in
Astroparticle Physics.
%
\bibitem{feldman} G.Feldman and R.Cousins, Phys. Rev. {\bf D57}
(1998)3873.
\end{thebibliography}
\end{document}